\documentclass[11pt]{article}
\usepackage{amsmath, amssymb, amsfonts}
\usepackage{graphicx,float}
\usepackage{natbib}
\usepackage{url, hyperref}
\usepackage{booktabs, multirow}
\usepackage{caption}
\usepackage{bbm}
\usepackage{lineno}
\usepackage{algpseudocode}
\usepackage{braket}
\usepackage{color}
\usepackage{geometry}
\usepackage{setspace}
\usepackage{algorithm}      

\title{Understanding How Network Geometry Influences Diffusion Processes in Complex Networks: 
A Focus on Cryptocurrency Blockchains and Critical Infrastructure Networks}

\author{
S M Mustaquim$^{1}$ \and 
Asim K. Dey$^{2}$\footnote{Corresponding author: \texttt{a.dey@ttu.edu}} \and 
Abhijit Mandal$^{1}$
}

\date{
$^{1}$Department of Mathematical Sciences, The University of Texas at El Paso, Texas, United States \\
$^{2}$Department of Mathematics and Statistics, Texas Tech University, Lubbock, Texas, United States
}

\begin{document}
\maketitle

\abstract{This study provides essential insights into how diffusion processes unfold in complex networks, with a focus on cryptocurrency blockchains and infrastructure networks. The structural properties of these networks, such as hub-dominated, heavy-tailed topology, network motifs, and node centrality, significantly influence diffusion speed and reach. Using epidemic diffusion models, specifically the Kertesz threshold model and the Susceptible-Infected (SI) model, we analyze key factors affecting diffusion dynamics. To assess the uncertainty in the fraction of infected nodes over time, we employ bootstrap confidence intervals, while Bayesian credible intervals are constructed to quantify parameter uncertainties in the SI models. Our findings reveal substantial variations across different network types, including Erd\H{o}s-R\'{e}nyi networks, Geometric Random Graphs, and Delaunay Triangulation networks, emphasizing the role of network architecture in failure propagation. We identify that network motifs are crucial in diffusion. We highlight that hub-dominated networks, which dominate blockchain ecosystems, provide resilience against random failures but remain vulnerable to targeted attacks, posing significant risks to network stability. Furthermore, centrality measures such as degree, betweenness, and clustering coefficient strongly influence the transmissibility of diffusion in both blockchain and critical infrastructure networks. 
}



\section{Introduction}
The study of diffusion processes in complex networks has gained significant attention in recent years due to its relevance in various domains, including epidemiology, social sciences, cybersecurity, power grid stability, financial systems and blockchain technology~\citep{broido2019scale,wu2020nowcasting,Albani_2024,Zhou_2021,ACEMOGLU2016536,greenwich33983,RIVA201911,dai2024,jrfm15020047,Eck_2021}. Complex networks, which encompass a diverse range of systems from financial transactions to power distribution infrastructures, exhibit distinct structural properties that directly influence the speed and reach of diffusion. Understanding how these processes unfold is crucial for designing more resilient and efficient networks, as well as mitigating vulnerabilities that could lead to cascading failures.

In blockchain networks, transaction propagation and validation depend heavily on network topology, with highly connected nodes (hubs) playing a critical role in ensuring efficiency and security~\citep{gervais2016is,conti2021survey,wang2022blockchain}. Similarly, critical infrastructure networks, e.g., electrical power grids, rely on stable energy distribution pathways to maintain functionality, and disruptions can result in widespread failures~\citep{rosas2018modeling,Liu_2021,Shan_2022}. The structural properties of these networks are key determinants of how diffusion processes unfold. Examining these characteristics provides valuable insight into the fundamental mechanisms that govern network behavior and resilience.

Existing research on diffusion processes in complex networks has extensively analyzed how information, failures, or innovations propagate through different network structures. \cite{Fornito2016,Costa2019,Schieber2023} study the effect of network distances, e.g., shortest paths and geodesical paths, on the diffusion process in networks. The influence of specific topological characteristics, such as modularity or degree heterogeneity of nodes, on diffusion processes,
 has been examined in~\cite{Motter2005,GomezGardenes2008,Payne2021,Peng2020}.
\cite{Wang2025} study the contact network uncertainty in epidemic inferences with approximate Bayesian computation. 
However, few approaches have systematically quantified the impact of specific higher-order network structures, such as 4-node motif concentrations~\citep{Menck_et_al2014, Schultz_et_al2014,Dey2019}, on diffusion speed across different network classes. Additionally, the integrated use of statistical inference and uncertainty quantification to assess the influence of structural features (e.g., motif frequency, centrality measures) and external covariates on network diffusion remains largely unexplored.

In this work, we address these gaps by correlating motif-level topology with diffusion speed in networks, while applying bootstrap and Bayesian techniques to quantify variability and statistical significance in the results. In particular, we thoroughly evaluate the effect of network structure and its topology on the diffusion process based on the Kertesz threshold diffusion  model~\citep{Kertesz_2015}. Using the bootstrap technique, we construct confidence intervals for the fraction of infected nodes over timesteps, providing a robust assessment of diffusion variability within the network. In addition, we explore the concept of network motifs to explore how the geometry of networks shapes the diffusion process. Finally, we utilize epidemiological compartmental models, specifically the Susceptible-Infected (SI) model~\citep{Hethcote2000,Aldous_2017}, along with Bayesian inference using the Markov Chain Monte Carlo (MCMC) algorithm, to evaluate the significant impact of network structure and external covariates on the diffusion process.

The remainder of this article is organized as follows. In Section~\ref{Sec:Methods}, we describe network diffusion models, different diffusion metrics, and statistical inferences. In Section~\ref{Sec:Simul}, we conduct simulation experiments to assess the effect of the network structure and its topology and geometry on the diffusion process. We analyze diffusions in cryptocurrency blockchain data in Section~\ref{Sec:cryptocurrency}. Section~\ref{Sec:power_grid} evaluates diffusion processes in critical infrastructure networks.  Finally, we conclude the paper with a discussion in Section~\ref{Sec:Discussion}.


\section{Diffusion in network}
\label{Sec:Methods}
We consider an undirected graph $G=(V, E, \omega)$ as a model for a complex network, where $V$ is the set of nodes, $E \subset \binom{V}{2}$ is the set of edges, and  $\omega \colon V \times V \mapsto \mathbb{R}_{\ge 0}$ is an edge weight function such that each edge $(u,v)\in E$ has a weight $\omega(\set{u,v}) = \omega_{uv}$. The adjacency matrix of the network with $n$ vertices is an $n \times n$ matrix where $A_{ij}=\omega_{ij}$ if $\set{i,j} \in E$,  and is 0 otherwise. Often, we are interested in a portion of the networks where $\omega$ is constant~\citep{halappanavar2015network,Young:NoN}.  In this case, it is more convenient to focus on the unweighted network, where  $\omega_{ij}=1$ for all $\set{i,j} \in E$, and is 0 otherwise. The total number of nodes in $G$ is $n = |V|$. 
The \textit{degree} $\deg(u)$ of a node $u$ is the  number of edges incident to $u$, i.e., for $u \in V$, $\deg{u} = \sum_{u \neq v} \mathbf{1}_{\set{u,v} \in E}$. For any integer $k \geq 0$, the \textit{degree distribution} $p(k)$ is the fraction of nodes over the whole network having degree $k$. This is, $p(k)$ is the probability that a randomly chosen node in the network has degree $k$. 
The \textit{average path length} and \textit{diameter} are two important parameters of a network. Let $\ell_{ij}$ be the number of edges (length) in the shortest path between two nodes $i$ and $j$. The average path length, denoted by $L$, is the mean of the shortest path length, averaged over all pairs of nodes:  $L= {\sum_{i\neq j} \ell_{ij}}/{n(n-1)} $. The diameter is the longest of all the calculated shortest paths in the network, i.e., $\text{diam}(G) = \max\limits_{ij} \{\ell_{ij}\}$. 

The \textit{clustering coefficient} measures the tendency of nodes in a network to form tightly connected groups by quantifying how often a node's neighbors are also connected. It reflects local cohesiveness, with higher values indicating well-connected communities and lower values suggesting a more fragmented structure. The clustering coefficient represents the probability that the adjacent nodes of a node are also connected. The \textit{betweenness centrality} captures how much a given node $u$ is in between other nodes. That is, the betweenness centrality measures the number of shortest paths between any couple of nodes in the graphs that pass through the target node $u$. This number is normalized by the total number of shortest paths existing between any couple of nodes of the graph. 
The \textit{giant component} of a network is a connected component that contains the vast majority of nodes~\citep{newman2018networks}. 

\subsection{Network diffusion model}
In a complex network, the failure of a single component can initiate a chain reaction, resulting in widespread cascading failures. In this paper, we use the words \textit{failure} and \textit{infection} interchangeably. The diffusion process is generally considered to be highly dependent on the network’s structure, topology, and geometry \citep{Akbarpour_2018, barabasi2016network, Bertagnolli_2021, Schieber_2023}. To analyze the dynamics of diffusion within a complex network and its various influencing factors, we employ epidemic diffusion models, specifically the \textit{Kertész threshold model}, on networks.

\cite{Kertesz_2015} proposed the Kertesz Threshold (KT) model as an extension of the widely recognized Watts threshold model \citep{Watts_2002} to study cascade dynamics in random networks. In the Watts threshold model, each node in the network monitors the states of its neighbors, which can be either 0 (susceptible) or 1 (infected). We use the terms `infection' and `adoption' interchangeably in the rest of the paper.  A node transitions to state 1 if a sufficient fraction of its neighbors are already in that state; otherwise, it remains in state 0. As a result, the Watts threshold model categorizes nodes into two types: Susceptible (S) and Adoption (A). The KT model expands on this by introducing a third category of nodes: Blocked (B), alongside Susceptible (S) and Adoption (A). Blocked nodes are resistant to external influence and do not change their state. Additionally, the KT model incorporates a spontaneous adopter rate ($\delta$) to account for external influences, allowing nodes to transition to state 1 independently of their neighbors. Thus, adoption in the KT model can occur either through neighbor influence or spontaneously due to endogenous effects. Algorithm~1 in Appendix~A summarizes the KT model.


\subsection{Diffusion metrics}

We evaluate how global network topology and higher order local geometry, e.g., the existence of motifs, affect the diffusion process. 
To evaluate the diffusion dynamics in networks under different topological and geometric properties we construct the following two metrics:

\begin{itemize}
    \item ~\textbf{Fraction of infected nodes:} The fraction of infected nodes at any given time \( t \), \( \eta(t) \), represents the proportion of the total nodes within the network that are infected at time $t$, and is defined as
    \begin{equation}\label{Eq:met1}
    \eta(t) = \frac{N_{\text{infected}}(t)}{N_{\text{total}}},
    \end{equation} 
    where \( N_{\text{infected}}(t) \) is the number of infected nodes at time \( t \), and \( N_{\text{total}} \) is the total number of nodes in the network.

    \item ~\textbf{Average speed of diffusion:} The average speed of diffusion, denoted as \( \nu \), quantifies the rate at which an infection reaches a steady state across the entire network. We can define the average speed of diffusion \( \nu \) as  
    \begin{equation}\label{Eq:met2}
       \nu = \frac{\eta(s) - \eta(0)}{t_s},
    \end{equation}
    where \( \eta(s) \) is the fraction of infected nodes at steady state, which equals 0.95 in all of our cases, indicating that all nodes that were not blocked become infected eventually, \( \eta(0) \) is the fraction of initially infected nodes at time \( t=0 \), \( t_s \) is the number of time steps required to reach the steady state.
        
\end{itemize}


\subsection{Motif concentration on diffusion speed} 
\begin{figure*}[!ht]
\centering
\includegraphics[width=0.35\textwidth]{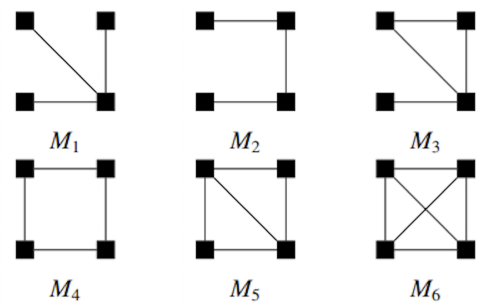}
\caption{All 4-node connected network motifs.}
\label{fig:motif}
\end{figure*}

A key characteristic of networks that significantly contributes to the study of complex systems is the concept of a \emph{network motif}. Traditional network metrics, such as node degree distribution, average path length, and diameter, primarily capture lower-order connectivity features at the level of individual nodes and edges. In contrast, network motifs serve as a measure of network geometry, emphasizing higher-order structures that reveal local interaction patterns within the network. These recurring subgraph structures provide deeper insights into the organizational principles and functional properties of complex networks.
Formally, a motif $G'=(V', E')$ is a typically small connected graph that occurs an abnormal number of times as an induced subgraph in $G$. Figure~\ref{fig:motif} illustrates all connected  4-node motifs ($M$) in a network~\citep{Milo2002}.

The occurrence of motifs in real-world networks is not random; instead, their presence or absence signifies that these substructures play a crucial role in shaping the network's functionality and structural organization~\citep{RosasCasals_CorominasMurtra2009,dey2020CJS}. We can evaluate the concentration, $C_i$, of an $n$-node motif of type $i$ as the ratio of its number of occurrences, $N_i$, to the total number of $n$-node motifs in the network. That is, 
$$C_i= \frac{N_i}{\sum_{i} N_i},$$
where $\sum_{i} N_i$ is the total number of $n$-node motifs~\citep{ahmed2016kais,Dey_2023}. 

In this study, we assess the impact of specific motifs on the diffusion process. Specifically, we analyze the correlation between the average diffusion speed (Eq.~\eqref{Eq:met2}) and the concentrations of various 4-node motifs. A positive correlation suggests that the presence of a particular motif accelerates diffusion, while a negative correlation indicates that the motif hinders the diffusion process.

To quantify the occurrence of motifs in each network, we employ a custom edge-disjoint motif detection algorithm based on the Weisfeiler-Lehman (WL) graph hashing \citep{Shervashidze2011}. The method focuses on connected 4-node subgraphs defined through six canonical motif templates. Candidate subgraphs are extracted from local neighborhoods and hashed using the WL kernel to generate an isomorphism-invariant signature, which is then matched against the predefined motif library. A subgraph is counted as a motif instance only if it is connected, matches one of the canonical templates, and shares no edges with any previously counted motif. This edge-disjointness constraint ensures that each counted edge participates in at most one motif, eliminating overlap-induced redundancy and yielding a conservative, non-redundant measure of motif prevalence. Algorithmic details of this procedure, including the generation of candidate subgraphs, WL hash computation, template matching, and edge-tracking for disjointness enforcement, are provided in Algorithm~3 in Appendix~C.


\subsection{Significance of structural node features in diffusion process}
In a complex network, diffusion of failure follows diverse pathways depending on several conditions, including the structure, topology, and geometry of the underlying network, the sensitivity to peer pressure, and the severity of the influence of external factors, e.g., natural hazards or intentional attacks. In a network to evaluate the effect of different factors in the diffusion process, we apply a Susceptible-Infectious (SI) model. The SI model has two diffusion states, i.e., the susceptible state and the infectious state. That is, in a SI model, individuals in a population are either infected or susceptible. The rate of the transition through the diffusion states is a function of both individual specific risk factors and the diffusion states of other individuals in the population at that time. In a SI model, the diffusion dynamics follow a Poisson process~\citep{Keeling_2005,Newman_2018,Angevaare_2022}. The model is defined on a \textit{continuous time}, where the time represents the length of time between events, i.e., inter-event periods, where the events are considered to be the state transition of diffusion.

In our context, the population represents a network, and individuals are network nodes. For each edge $(u,v) \in E$, if at some time one node ($u$ or $v$) becomes infected while the other is susceptible, then the other will later become infected with some transmission probability. 
That is, the edges in the network represent potential transmission routes through which diffusion can spread between nodes in the network~\citep{Aldous_2017,Spricer_2019}.

Under the SI model framework, the transition rate for a \textit{susceptible} node, $i$, to the \textit{infectious} state during the $t^{th}$ time period can be defined as
\begin{equation}\label{Eq:SI_ILM}
\lambda_{SI}(i, t) =  \psi_S(i) \sum_{j \in I(t)} \psi_T(j) \kappa(i, j) + \epsilon(i),~\text{for}~ i \in \mathcal{S}_{(t)}, j \in \mathcal{I}_{(t)},
\end{equation}
where $\mathcal{S}_{(t)}, \mathcal{I}_{(t)}$ are the set of susceptible, and infectious nodes during the $t^{th}$ time period, respectively. 
Each individual in the population is considered to be in one of the two diffusion states at any $t^{th}$ time period, i.e., $N=\mathcal{S}_{(t)} +\mathcal{I}_{(t)}$, where $N$ is the total number of nodes (individuals). The function $\psi_S(i)$ is a susceptibility function that represents the risk factors associated with the susceptible node $i$ contracting the spread, and $\psi_T(j)$ is a transmissibility function that represents the risk of infection transmission from the $j$th node. The function $\kappa(i,j)$ is an infection kernel function that describes the shared risk factors between pairs of infectious and susceptible nodes, $i$ and $j$, respectively. The $\epsilon(i)$ is a sparks function function that represents the additional risk factors associated with exposure to the the $i^{th}$ node that the model otherwise fails to explain. Typically, this refers to exposure from a non-specified source outside of the observed population~\citep{Deardon_2010,Pokharel_2016}.

The susceptibility and transmissibility functions, $\psi_S(i)$ and $\psi_T(j)$, respectively, can incorporate any node-level covariates of interest. We can define $\psi_S(i)$ in Eq.~\eqref{Eq:SI_ILM} as a linear function of the covariates as 
\begin{equation}\label{Eq:SI_ILM_sus}
\psi_S(i)= \omega_0 + \omega_1 X_1(i) + \ldots + \omega_{p_1} X_{p_1}(i) ,
\end{equation}
where $X_1(i), \ldots,  X_{p_1} (i)$ are the $p_1$ covariates associated with susceptible node $i$ along with susceptibility parameters $\omega_1, \ldots, \omega_{p_1}$. Similarly, $\psi_T(j)$ in Eq.~\eqref{Eq:SI_ILM} can be treated as a linear function of the covariates as 
\begin{equation}\label{Eq:SI_ILM_Trans}
\psi_T(j)= \phi_1 X_1(j) + \ldots + \phi_{p_2} X_{p_2}(j) ,
\end{equation}
where $X_1(j), \ldots,  X_{p_2}(j)$ are the $p_2$ covariates associated with susceptible infectious $j$ along with transmissibility parameters $\phi_1, \ldots, \phi_{p_2}$. The kernel function $\kappa(i,j)$ can be defined as 
\begin{equation}\label{Eq:connectivity}
\kappa(i,j)=  \alpha d^{-\gamma}_{ij} + \theta_1 {C}_{ij}^{(1)} + \ldots + \theta_{p_3}{C}_{ij}^{(p_3)},
\end{equation}
where $d_{ij}$ is the distance between nodes $i$ and $j$, $\alpha$ represents a scaling parameter, and $\gamma$ is the exponent of the distance. ${C}_{ij}^{(.)}$s represent network matrices. The parameters $\theta_{(.)}$s represent the effect of each of the $p_3$ network matrices on transmission risk. ${C}_{ij}^{(.)}$ can be either a categorical variable or a quantitative variable (especially, in a weighted network). In a categorical variable  ${C}_{ij}^{(.)}=1$, if $i$ and $j$ belong to one group, and 0, otherwise. In a quantitative variable, ${C}_{ij}^{(.)} \in \mathcal{R}$. We set $\kappa(i, j)$ equals zero if no connection exists~\citep{Vineetha_2020,Angevaare_2022}.

The $\lambda_{SI}(i, t)$ describes the spread of diffusion in a SI model on a network. As a Poisson process, the inter-event periods follow exponential distributions. The likelihood function for the associated parameters, $\Theta=(\omega_0, \omega_1,\ldots, \omega_{p_1}, \phi_1,\ldots, \phi_{p_2}, \alpha, \gamma, \theta_1,\ldots,\theta_{p_3})^T$, is the product of likelihoods at time periods indexed by $t = 1,\ldots, T-1$, where $T$ is the total number of events that have occurred. We can write the likelihood function as
 \begin{equation}
    L(\Theta) = \prod_{t=1}^{T-1} \lambda(t) v(t) \exp\{-v(t)\Delta_t\},
 \end{equation}
where $v(t)=\sum_{i\in \mathcal{S}_{(t)}} \lambda_{SI}(i, t)$, and $\lambda(t) = {\lambda_{SI}(i, t)}/{v(t)}$, if $i \in \mathcal{S}_{(t)} \cap \mathcal{I}_{(t+1)}$, i.e., $i$ transitioned from susceptible to infected. The length of time since the beginning of the $t^{th}$ time period is denoted as $\Delta_t$. 
The parameters of the SI model in the network are estimated using Bayesian inference via the Markov chain Monte Carlo (MCMC) method, particularly, the Metropolis-Hastings (M-H) algorithm. The M-H algorithm generates a Markov chain consisting
of a sequence of samples within the parameter space, where the distribution of these samples converges to the posterior distribution of the parameters. To begin the chain, we start with an initial guess $\Theta_0$ of the parameters. After initialization, we propose new samples using a transition kernel. For a symmetric transition kernel, the proposal at iteration $w$, denoted $\Theta_w$, is accepted with probability:
\begin{equation}
\pi(\Theta_w, \Theta_{w-1}) = P(\Theta_{w-1} \rightarrow \Theta_w) = \min\left(1, \frac{L(\Theta_w) \, P(\Theta_w)}{L(\Theta_{w-1}) \, P(\Theta_{w-1})} \right),
\end{equation}
where $L(\Theta)$ is the likelihood function and $P(\Theta)$ is the prior distribution. If the proposal $\Theta_w$ is rejected, we retain the current state, i.e., $\Theta_w = \Theta_{w-1}$~\citep{Hastings_1970,
robert2013monte}. We perform the estimation procedure using the Julia package \textit{Pathogen.jl} using weekly informative priors~\citep{Angevaare_2022}.

\section{Simulation studies}
\label{Sec:Simul}

\subsection{Diffusion speed in classical random graph models}
In this section, our goal is to evaluate the influence of network structure and its topology on the diffusion process. In particular, we want to assess the speed of diffusion with uncertainty quantification in different random graph models, e.g., Erd\H{o}s-R\'{e}nyi graph, geometric random graph, and network based on Delaunay triangulation, through an extensive simulation study. Below, we define the three random graph models used in our analysis.

\begin{itemize}
   \item ~\textbf{Erd\H{o}s-R\'{e}nyi (ER) graphs:} Erd\H{o}s-R\'{e}nyi graphs are fundamental in network theory and are widely used to model various phenomena, including social and computer networks~\citep{ErdosRenyi1959,PhysRevE.85.056109, 7952848}. 
   The Erd\H{o}s-R\'enyi graph, denoted as \( G(n, p) \), is a model of a random graph consisting of \( n \) vertices where an edge connects each distinct pair of vertices with a probability \( p \), independently of other edges. Formally, the graph \( G \) is defined by the vertex set \( V = \{1, 2, \dots, n\} \) and an edge set \( E \), where each edge \( \{i, j\} \) for \( i \neq j \) is included in \( E \) with probability \( p \). The resulting graph structure \( G = (V, E) \) exhibits key statistical properties: the expected number of edges is \( E(\lvert E \rvert) = p \binom{n}{2} \), and the degree of each vertex follows to a binomial distribution \( \text{Bin}(n-1, p) \)~\citep{erdos1960}.

\item ~\textbf{Geometric Random Graph (GRG):} 
Geometric random graphs are a fundamental model for networks constrained by spatial factors, making them valuable in applications such as cyber-physical networks (e.g., wireless communication, sensor networks, and transportation networks) and biological systems. In a geometric random graph, \(n\) vertices are randomly placed in a metric space, typically a unit square or unit disk. Two vertices are connected by an edge if their Euclidean distance is within a specified threshold \(r\)~(\citealp{Dall2002,penrose2003,Barthelemy2011}). This model effectively captures spatial dependencies and clustering properties that are not present in purely random graph models, such as Erd\H{o}s-R\'{e}nyi graphs. The expected degree of a node in a geometric random graph depends on the radius \(r\) and the node density, both of which influence the network's connectivity and diffusion dynamics.

\item  ~\textbf{Delaunay Triangulation (DT) network:} 
A Delaunay Triangulation (DT) network is a graph-based structure derived from the Delaunay triangulation, where the points in the network are connected based on the geometric properties of the DT. In this network, edges are formed between points that share a triangle in the triangulation, ensuring that the network captures the spatial relationships and local connectivity inherent in the data. Formally, a DT,  
\( \mathcal{D}(V) \), of a finite set of points \( V = \{v_1, v_2, \ldots, v_n\} \) in the Euclidean plane \( \mathbb{R}^2 \) is a triangulation such that no point in \( V \) lies inside the circumcircle of any triangle in the triangulation, where for each triangle \( \triangle abc \) in \( \mathcal{D}(V) \), the circumcircle \( C \) of \( \triangle abc \) satisfies \( C \cap V = \{a, b, c\} \). The DT maximizes the minimum angle among all possible triangulations of \( V \), avoiding degenerate skinny triangles. It can be constructed by ensuring that each edge \( (v_i, v_j) \) belongs to \( \mathcal{D}(V) \) if and only if there exists a circle passing through \( v_i \) and \( v_j \) that contains no other points from \( V \) in its interior. This triangulation is the geometric dual of the Voronoi diagram, and it is unique when no four points in \( V \) are cocircular~\citep{preparata1985,Fortune1992}. DT networks are frequently applied in fields such as wireless sensor networks, power grid networks, and spatial networks, as they provide a natural way to model the connectivity of a set of points while optimizing for properties like minimal edge lengths and avoiding long, thin connections~\citep{Chew1989,Aurenhammer1991}.

\end{itemize}

We apply the KT model on ER, GRG, and DT networks, and evaluate the fraction of nodes infected over time. For each network type, we generate 1,000 instances, where each instance is created using specific parameters: 300 nodes with a connection probability of 0.02 for ER graphs, 300 nodes with a connection radius of 0.08 for GRG, and 300 normally distributed points for the DT graphs. These parameters were chosen such that the expected number of edges is approximately the same (around 900) across all three network types, ensuring that the average degree is similar (approximately 6) and that differences in the diffusion dynamics arise primarily from the network topology rather than density. Specifically, for the ER graphs, the expected number of edges is given by $p \cdot \frac{n(n-1)}{2}$, which yields $\approx 897$ for $n=300$ and $p=0.02$. For GRG graphs, the expected number of edges is approximated by $\frac{\pi r^2 n(n-1)}{2}$, giving $\approx 900$ when $r = 0.08$. For DT graphs, the number of edges is deterministic for planar graphs, with $3n - 6 = 894$ for $n=300$. A visual representation of these three random graph networks is shown in Online Supplementary Material Figure~S1.

The parameters of the KT model are experimentally determined as follows: the initial fraction of infected nodes ($\eta$) is set to 0.01, the number of iterations ($k$) is fixed at 500, the adopter rate ($\delta$) is 0.001, the proportion of blocked nodes ($\beta$) is 0.05, and the node threshold ($\tau$) is assigned values of 0.30 and 0.40. We calculate the mean fraction of infected nodes at each timestep for each graph type. However, the speed of diffusion is influenced by the centrality of the initially infected nodes. Specifically, diffusion spreads more rapidly when the initial infected nodes have high centrality compared to cases where they have moderate or low centrality~\citep{Schieber_2023}. We select the initial infected nodes based on their structural importance in the network, specifically using degree and betweenness centrality measures. For each network instance, we identify the top 1\% of nodes with the highest degree and the top 1\% with the highest betweenness centrality, and use them as initial infection sources in separate simulation scenarios.

\begin{figure*}[!htbp]
    \centering
    \includegraphics[width=0.49\textwidth]{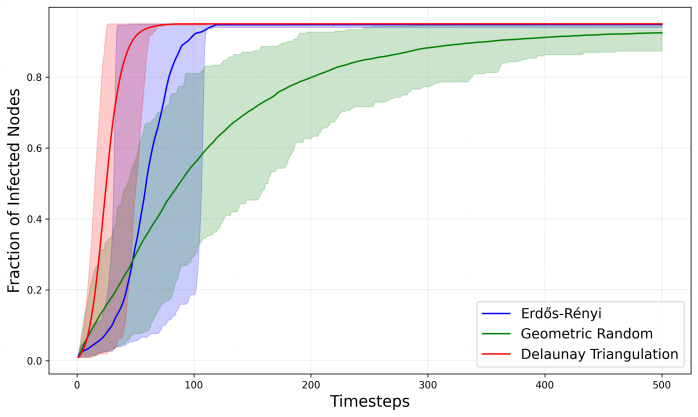}
     \includegraphics[width=0.49\textwidth]{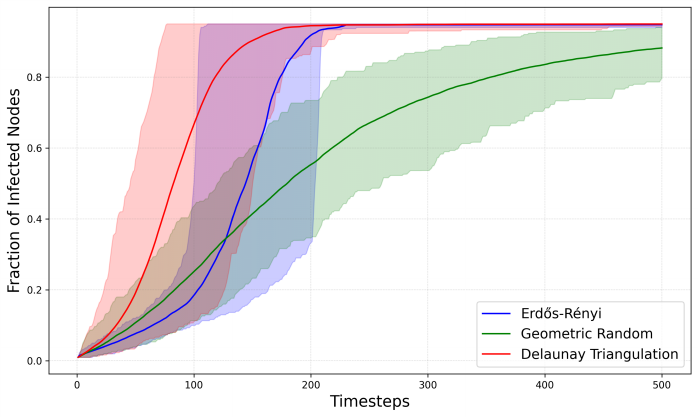}
   \caption{The effect of network structure/topology on the diffusion process. The left panel represents diffusion starting with the top three nodes with the highest degree for $\tau  = 0.30$ and the right panel represents the diffusion for  $\tau  = 0.40$, with their corresponding $95\%$ confidence intervals.  }
    \label{fig:three-models}
\end{figure*}

Figure~\ref{fig:three-models} illustrates the fraction of infected nodes over time for the three random graph models, with 95\% confidence intervals, where the initially infected nodes are chosen based on the highest degree centrality. 
The 95\% confidence intervals have been determined from 1,000 independent simulation runs for each graph type and timestep. Specifically, for each timestep $t$, we calculated the mean fraction of infected nodes along with the corresponding bounds using the empirical $2.5^{th}$ and $97.5^{th}$ percentiles derived from the simulation runs. A detailed description of the full procedure can be found in Algorithm 2 in Appendix B. The results indicate that nodes in DT networks are infected at the fastest rate, while those in GRG networks experience significantly slower infection. This suggests that among the three random graph models, the DT network is the most vulnerable to diffusion processes, whereas the GRG network is the most resilient. A similar diffusion pattern is observed when the initial infected nodes are selected based on the highest betweenness centrality (see Online Supplementary Material Figure~S2).

\subsection{Motif concentration on diffusion speed} 

In this section, we conduct a comprehensive simulation study using Stochastic Block Model (SBM) networks to analyze how specific 4-node motifs influence the diffusion process in the network. A SBM network has $k$ number of equal-sized communities (blocks), and the intra-community and inter-community connectivity for each community can be represented with a $k\times k$ symmetric probability matrix $P$. The diagonals of $P$ correspond to the probability of an edge within blocks, and the off-diagonals correspond to the probability of an edge between blocks~\citep{Karrer2011,Pavlovi2014StochasticBO,Dey_2022_B,Peixoto2018,Ng2021}. We denote the within and between community connectivity probability by $p_{w}$ and $p_{b}$, respectively. 
We consider a SBM network consisting of \( n = 600 \) nodes partitioned into \( k = 3 \) communities, each containing 200 nodes. 
We can write the probability matrix $P$ for $k=3$, $p_{w}=0.030$, and $p_{b}=0.001$ as 
\begin{equation}\label{Eq:BPM}\nonumber
P = 
\begin{pmatrix}
0.030 & 0.001 & 0.001  \\
0.001 & 0.030 & 0.001 \\
0.001 & 0.001 & 0.030 \\
\end{pmatrix}.
\end{equation}
Online Supplementary Material Figure~S3 represents an SBM network instance with its expectation adjacency matrix. To explore variations in connectivity, in this experiment, the off-diagonal probabilities in the block probability matrix are systematically adjusted across 10 different values, ranging from 0.001 to 0.01, while the diagonal probabilities remain fixed at 0.03. Each configuration undergoes 1,000 independent simulation runs, extending up to 500 time steps or until all nodes are infected. During each simulation, we determine the average diffusion speed using Eq.~\eqref{Eq:met2}, monitor the occurrence of specific 4-node motifs, and compute their concentrations. Finally, we compute the correlations between diffusion speed and  concentrations of different 4-node motifs ($M_1$ to $M_3$). The frequencies of $M_4$, $M_5$, and $M_6$ motifs are very low in each simulation; therefore, we exclude them from the experiment.

Table~\ref{tab:correlation_matrix} presents the average correlation coefficients along with their standard deviations, calculated from 1,000 simulations. 
The mean correlation values indicate the strength and direction of the relationship between motif concentrations and diffusion speed, while the standard deviation reflects the variability across simulations. We find that motifs $M_1$ and $M_2$ exhibit very high positive correlations (0.9447 and 0.8345, respectively), suggesting that networks with higher concentrations of star ($M_1$) and line-type ($M_2$) motifs tend to experience faster diffusion. That is, the structural properties of $M_1$ and $M_2$ facilitate the spread of diffusion within the network. Similarly, $M_3$ has a modest positive correlation (0.4776) with a higher standard deviation (0.2554), but its effect appears less pronounced compared to $M_1$ and $M_2$. 

\begin{table*}[!ht]
\centering
\caption{Correlations between average diffusion speed and 4-node motif concentrations.}
\label{tab:correlation_matrix}
\begin{tabular}{@{}lcc@{}}
\toprule
Motif    & Mean         & Standard deviation \\ \midrule
$M_1$       & 0.9447   & 0.0385     \\
$M_2$       & 0.8345   & 0.0860     \\
$M_3$       & 0.4776   & 0.2554 \\
\bottomrule
\end{tabular}
\end{table*}



\section{Diffusion in cryptocurrency blockchain network}
\label{Sec:cryptocurrency}

The diffusion process in cryptocurrency blockchains refers to how information, such as transactions or communication between agents (i.e., miners and users), spreads across the network. This process is crucial for maintaining the integrity and efficiency of the blockchain~\citep{jrfm15020047,Yu2020,Vasudevan_2024}. In this section, we evaluate the dynamics of the diffusion process in the Ethereum transaction network. The Ethereum transaction data are collected from the Ethereum mainnet using Infura, a blockchain node infrastructure service that offers instant and scalable API access to the Ethereum network~\citep{Infura_2024}. The Ethereum ETL tool~\citep{ethereum2etl} is utilized to convert blockchain data into accessible formats, specifically extracting traces from blocks 10,000,000 to 10,001,000. In this study, we focus on transactions filtered within the timeframe from 2020-05-04 14:48:40 (UTC) to 2020-05-04 14:58:40 (UTC). We have constructed a weighted network from this Ethereum transaction data, where nodes represent accounts, and an edge exists between two nodes if a transaction occurs between them. The edge weights correspond to the transaction volume. The resulting network is highly sparse, with many disconnected nodes. Therefore, we only consider the giant component of the network, which consists of 627 nodes and 666 edges. 

\begin{figure}[htbp]
    \centering
    \includegraphics[width=0.48\textwidth]{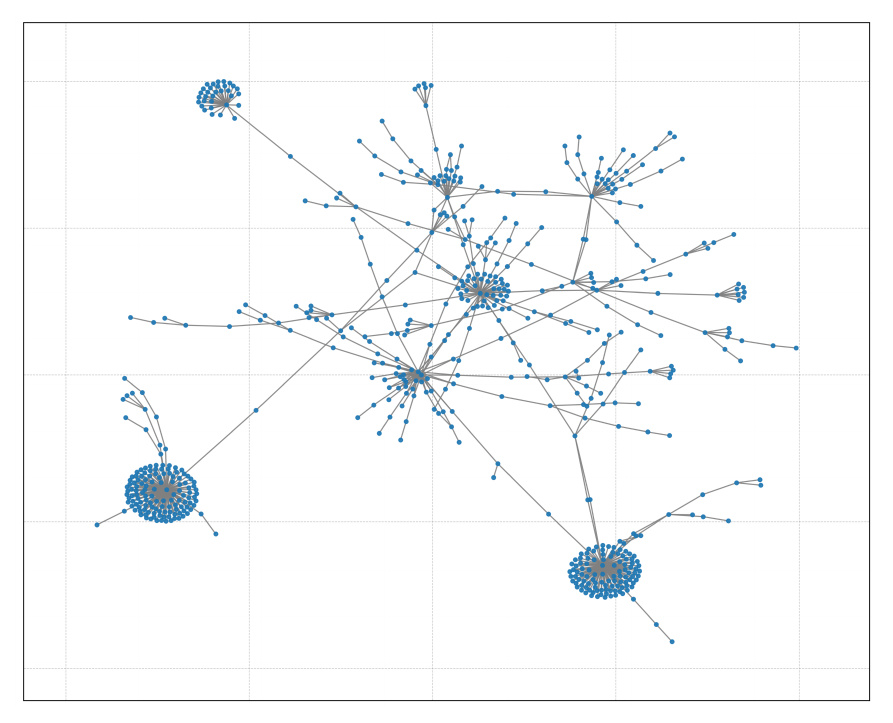}
    \includegraphics[width=.48\textwidth]{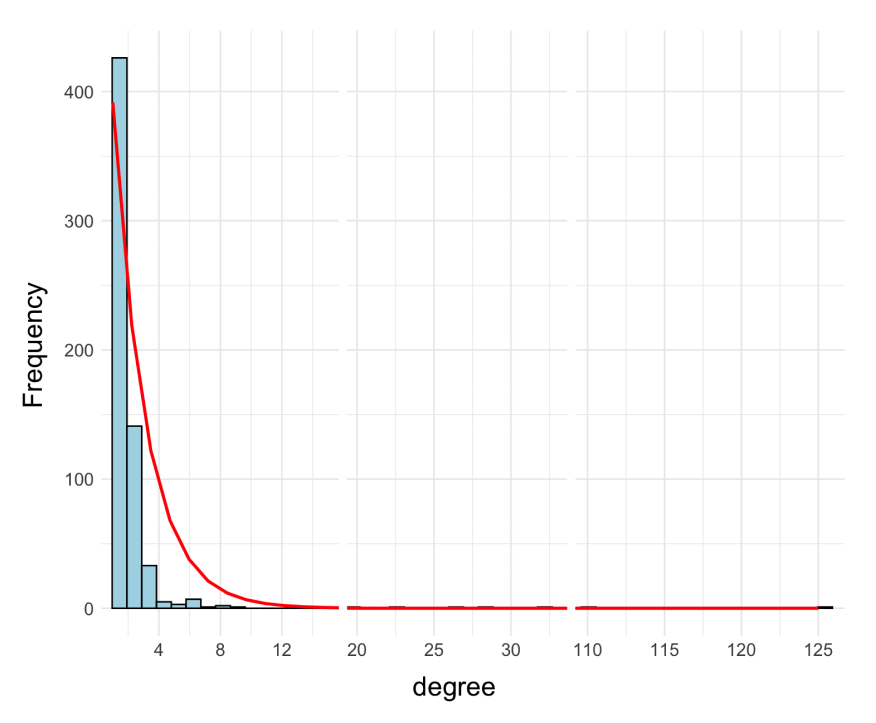}
    \caption{Ethereum transaction network (left) and its degree distribution with fitted exponential curve (right).}
    \label{fig:ethereum_analysis}
\end{figure}

We observe that the constructed Ethereum transaction network (left panel of Figure~\ref{fig:ethereum_analysis}) exhibits a right-skewed, hub-dominated degree distribution~\citep{DeCollibus2021,Serena2022}, characterized by a small number of highly connected hub nodes while the majority of nodes have very few connections. As illustrated in the right panel of Figure~\ref{fig:ethereum_analysis}, this structure means that a few dominant hubs play a crucial role in maintaining the overall connectivity of the network.

\subsection{KT model in Ethereum transaction network}

In this section, we apply the KT model to the Ethereum transaction network. The KT model emulates the dynamics of cascading failure of nodes in the network. We have observed that the Ethereum transaction network exhibits a hub-dominated, heavy-tailed degree distribution \citep{DeCollibus2021,Serena2022}, which makes the network highly resilient to random failures but particularly vulnerable to targeted attacks on these highly connected nodes \citep{Albert2000}. To assess the diffusion dynamics under these scenarios, the initial infected nodes in the KT model are selected using three strategies: (i) randomly, (ii) based on the highest degree centrality, and (iii) based on the highest betweenness centrality. The remaining parameters of the KT model are determined experimentally as follows: the fraction of initially infected nodes ($n$) is set to 0.01, the number of iterations ($k$) is 500, the adopter rate ($\delta$) is 0.001, the proportion of blocked nodes ($\beta$) is 0.05, and the node threshold ($\tau$) is 0.40.

\begin{figure}[!t]
		\centering%
		\begin{tabular}{c} 
			\includegraphics[width=0.85\textwidth]{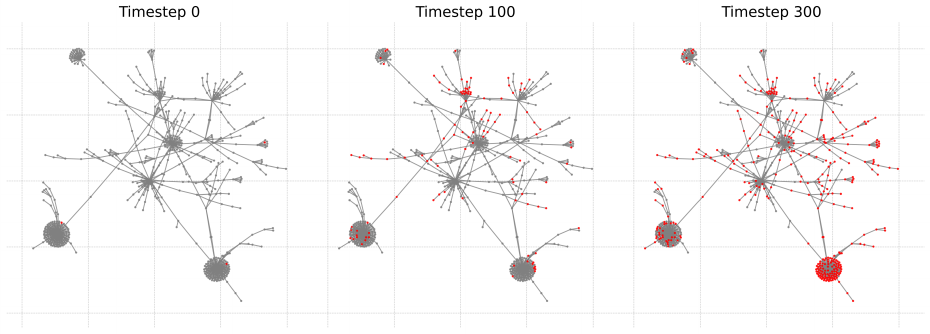}\\ 
           (a) Initial infected nodes are selected randomly.  \vspace{.2in}\\
			\includegraphics[width=0.85\textwidth]{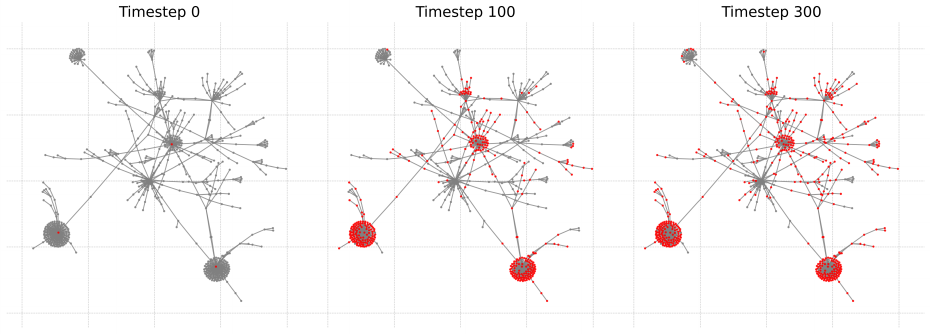}\\
            (b) Initial infected nodes are selected based on the highest degree centrality. \vspace{.2in}\\
            \includegraphics[width=0.85\textwidth]{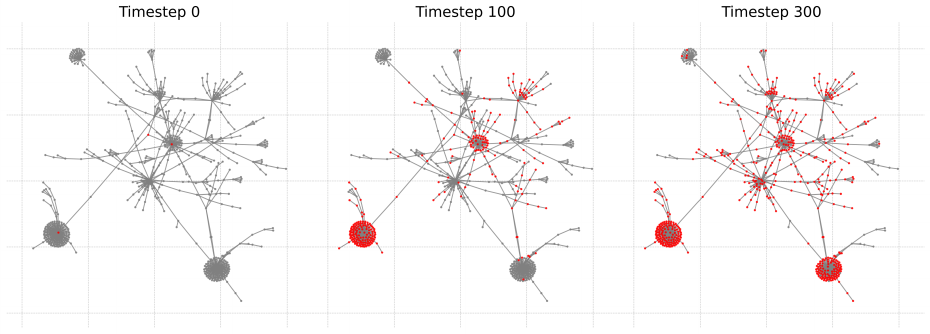}\\
            (c) Initial infected nodes are selected based on the highest betweenness centrality.
			 \end{tabular}%
		\caption{Spread of node failures in the Ethereum network under diffusion based on the KT model, where gray nodes represent uninfected nodes and red nodes represent infected nodes.}
		\label{fig:KT_ETH}
	\end{figure}

Figure~\ref{fig:KT_ETH} illustrates the diffusion of node failures in the Ethereum transaction network using the KT model, where the initial infected nodes are chosen either randomly or based on the highest centrality measures. The results indicate that diffusion progresses significantly slower when the initial infections are randomly assigned compared to when nodes with the highest degree or betweenness centrality are targeted.

\begin{figure*}[!ht]
    \centering
        \includegraphics[width=0.50\textwidth]{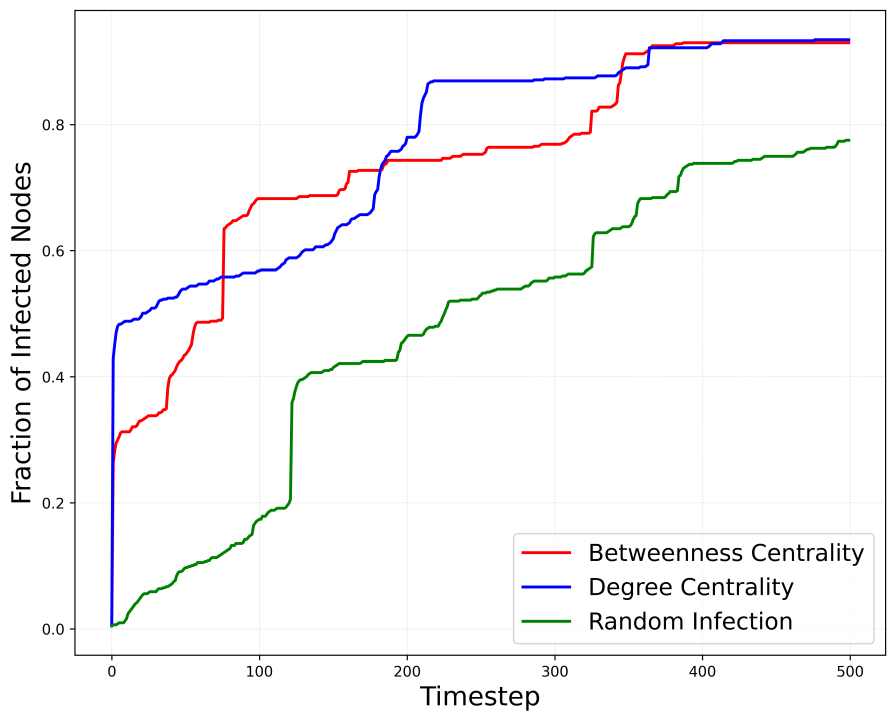}
    \caption{Comparison of fraction of infected nodes based on the KT model over time steps in Ethereum network under three different initial infection scenarios. } 
    \label{fig:comparison_ETH}
\end{figure*}

Now, we compute the fraction of infected nodes at different time steps $t$ based on Eq.~\eqref{Eq:met1}. 
Figure~\ref{fig:comparison_ETH} illustrates the fraction of infected nodes in the Ethereum transaction network over time for three different initial infected node selection strategies. The results show that when the initial infected nodes are selected randomly, the fraction of infected nodes remains significantly lower compared to cases where selection is based on the highest degree or betweenness centrality. Additionally, between timesteps 1 and 100, the fraction of infected nodes is higher when the initial infections are based on degree centrality than when based on betweenness centrality. This trend reverses between timesteps 101 and 200, with betweenness centrality leading to a higher infection rate, before shifting back again between timesteps 201 and 400.

We also evaluate the differences in diffusion rates across the three initial infected node selection scenarios in the Ethereum transaction network using the nonparametric $k$-sample Anderson-Darling test. The $k$-sample Anderson-Darling test is the generalization of the two-sample Anderson-Darling test~\citep{Pettitt_1976}. The test evaluates the hypothesis that $k$ independent samples with sample sizes $n_1, n_2, \ldots, n_k$ are from a common unspecified distribution~\citep{Scholz_1987,neuhauser2011nonparametric}. The test result shows significant differences among the diffusion rates in the three initial infected node selection scenarios ($p$-value $< 0.001$).

The outputs of the KT model highlight that the Ethereum transaction network is resilient to random attacks due to the large number of low-degree nodes. However, a targeted attack on key hubs (such as large exchanges or mining pools) could severely impact transaction propagation and network stability.


\subsection{SI model on Ethereum blockchain network}

To evaluate the effect of network topology in a diffusion process, we apply the SI model to the Ethereum transaction network. Under the SI model, the infection (failure) transition rate for a susceptible blockchain account $i$ to the infectious
state during the $t^{th}$ time period is defined based on Eq.~\eqref{Eq:SI_ILM}. In this case, we define the susceptibility function in Eq.~\eqref{Eq:SI_ILM_sus} as $\psi_S(i) = 1.0$. That is, we uniformly assign a constant susceptibility value of 1.0 to every Ethereum blockchain account $i$ in the network. The transmissibility function $\psi_T(j)$ quantifies the overall transmission risk of a susceptible Ethereum blockchain account $j$ by incorporating normalized centrality measures and clustering coefficients as
\begin{equation}\nonumber
\psi_T(j)= \phi_1 X_1(j) + \phi_{2} X_{2}(j) + \phi_{3} X_{3}(j),
\end{equation}
where $X_1$ represents normalized degree centrality, $X_{2}$ represents normalized betweenness centrality, and $X_{3}$ normalized clustering coefficient of the account $j$. The normalization is done by dividing their respective maximum values in the network.

To evaluate connections between blockchain accounts, the connectivity kernel $\kappa(i, j)$ in Eq.~\eqref{Eq:connectivity} is employed. This function utilizes a power-law relationship that depends on the shortest path distance $d_{ij}$ between two connected accounts $i$ and $j$. The $\kappa(i, j)$ also includes a normalized transaction value between the accounts, $C_{ij}$, as an additional covariate. We can write the connectivity kernel as
\[
\kappa(i, j) =
\begin{cases} 
\alpha d_{ij}^{-\gamma}  + \theta C_{ij}, & \text{if there is a connection between } i \text{ and } j, \\
0, & \text{otherwise},
\end{cases}
\]
 The sparks function  assigns a constant risk value, denoted as $\epsilon(i) = \zeta$ (Eq.~\eqref{Eq:SI_ILM}), to each Ethereum account. We assign weakly informative priors based on the following distributions

\[
\zeta \sim \text{Exp}(0.0001), \quad \alpha \sim \mathcal{U}(1, 2),
\]
\[
\gamma \sim \mathcal{U}(0.01, 5), \quad \theta \sim \mathcal{U}(1000, 100000),
\]
\[
\phi_1 \sim \mathcal{U}(0, 10), \quad \phi_2 \sim \mathcal{U}(0, 10).
\]
\[
\phi_3 \sim \mathcal{U}(0, 10).
\]

The prior parameter values were chosen to generate an informative diffusion process~\citep{Pokharel_2016,Angevaare_2022}, ensuring that the diffusion would neither fade away too quickly nor spread through the entire population too rapidly. 
We use  Julia library Pathogen.jl to fit the model. The Pathogen.jl uses the Metropolis-Hastings MCMC algorithm to produce model estimates for the SI model in the Ethereum blockchain network. The number of iterations used was 200,000, with a burn-in period of 20,000. The computation took approximately 48 hours on a workstation with 32 GB of RAM and a 4.40 GHz Intel Core i7 processor.

Table~\ref{Bayes_est2} presents the posterior mean estimates of the SI model parameters along with their 95\% credible intervals for the Ethereum blockchain network. A parameter is considered statistically significant if its 95\% credible interval does not include zero. In addition, a narrow credible interval of the SI model parameters indicates strong evidence about a parameter, while a wide credible interval suggests high uncertainty. The results indicate that the estimated coefficient for external source risk ($\hat{\zeta}$) is very low (0.0001), suggesting that infections from external sources are rare. Additionally, transaction value between nodes exerts significant infectious pressure ($\hat{\theta}=30041.2$). We also observe that degree centrality, betweenness centrality, and clustering coefficient play significant roles in determining transmissibility. Notably, the clustering coefficient has a substantially greater impact on the diffusion process compared to degree and betweenness centrality. The MCMC diagnostic plots (Online Supplementary Material Figure~S5) confirm the convergence of the model parameters.

\begin{table}[h]
\centering
\caption{Parameter estimates with 95\% credible intervals for Ethereum Transaction Network.}
\label{Bayes_est2} 
\begin{tabular}{cccc}
    \hline
    Parameter & Mean &  95\% credible interval\\
    \hline
    $\zeta$ & 0.0001 & (3.804$\times10^{-5}$, 0.0002) \\
    $\alpha$   & 0.0616   & (0.0107, 0.1872)          \\
    $\gamma$   & 1.5190     &  (1.0148, 1.9668)              \\
    $\theta$   & 30041.2     &  (9391.13, 63724.2)             \\
    $\phi_1$ & 1.9022     &  (0.6492, 4.4153)            \\
    $\phi_2$ & 1.0475     &  (0.1119, 3.0667)            \\
    $\phi_3$ & 4.5457     &  (0.0034, 9.7166)          \\
    \hline
\end{tabular}
\end{table}


The investigation into diffusion dynamics within cryptocurrency blockchain networks, particularly focusing on Ethereum, reveals significant societal implications concerning digital infrastructure resilience, financial equity, and systemic risk. Our analysis shows that the Ethereum network exhibits a hub-dominated topology, marked by a select group of highly interconnected hubs. These critical nodes are essential for maintaining network cohesion and ensuring the efficient propagation of transactions. While this structure enables robust performance under random conditions, it simultaneously renders the network vulnerable to targeted attacks. The failure or compromise of these key hubs, such as major exchanges or mining pools, could disrupt not only the technical operation of the network but also diminish trust in decentralized financial systems, ultimately undermining public confidence in blockchain technology as a viable alternative to traditional finance. The uneven distribution of influence within the network, as reflected by metrics like degree and betweenness centrality, may deepen existing digital inequalities. Entities that control these high-centrality nodes gain disproportionate visibility and power over transaction flows, challenging the egalitarian ideals often associated with decentralized systems. This can have concrete effects on market behavior, access to liquidity, and participation in decentralized applications, especially for underrepresented or marginalized users. Furthermore, our study highlights that these central nodes are also primary drivers of diffusion under both the KT and SI models, indicating that their influence extends beyond transaction activity to shape the speed and patterns of systemic disruptions and information propagation within the network.



\section{Diffusion in critical infrastructure networks}
\label{Sec:power_grid}

The diffusion process in critical infrastructure networks refers to the spread of failures, information, or disruptions across essential systems such as power grids, transportation, water supply, and telecommunications networks. The component failures in such networks -- whether caused by cyberattacks, natural disasters, or technical issues -- can rapidly cascade, leading to large-scale disruptions. In this study, we examine diffusion dynamics in critical infrastructure networks, with a particular focus on electricity transmission systems of Italy, France, and Germany. The data are obtained from the Union for the Coordination of the Transmission of Electricity. In an electricity transmission network nodes represent power stations/substations, and edges represent physical transmission lines connecting two nodes. The numbers of nodes, edges, and other topological properties of the three power grid networks are listed in Table~\ref{tab:EU}. 
 
\begin{table*}[!ht]
	\caption{Properties of the three European power grid networks.}
	\label{tab:EU}	
	\centering
	\begin{tabular}{l*{6}{c}r} \hline
		Power grid & Nodes & 	Edges & Average path length & Diameter \\
		\hline
		Germany   & 445 & 567 & 11.7510  & 30 \\		
		Italy & 273 & 375  & 9.7420 & 28 \\
		
		France & 677 & 913 & 9.5920  & 26  \\
		\hline
	\end{tabular}
\end{table*}

\subsection{KT model in European power grids}

In this section, we employ the KT model to analyze three real power grid networks, simulating the diffusion dynamics of node failures within these systems. The initial infected nodes are selected using two targeted attack strategies: (i) based on the highest degree centrality and (ii) based on the highest betweenness centrality. The remaining KT model parameters are experimentally set as follows: the fraction of initially infected nodes ($\eta$) is 0.01, the number of iterations ($k$) is 200, the adopter rate ($\delta$) is 0.001, the proportion of blocked nodes ($\beta$) is 0.05, and the node threshold ($\tau$) is 0.30.

\begin{figure}[!t]
		\centering%
		\begin{tabular}{c} 
			\includegraphics[width=0.85\textwidth]{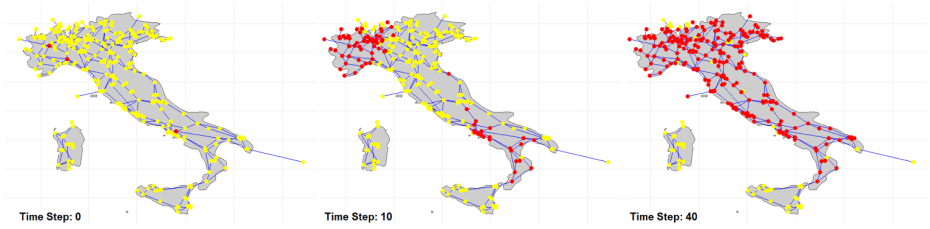}\\ 
           (a) Italian power grid  \vspace{.2in}\\
			\includegraphics[width=0.85\textwidth]{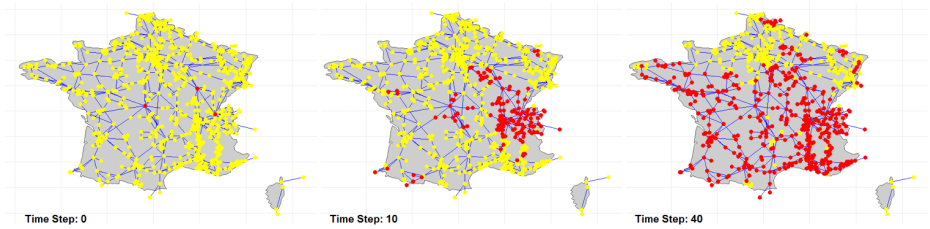}\\
            (b) French power grid \vspace{.2in}\\
            \includegraphics[width=0.85\textwidth]{
            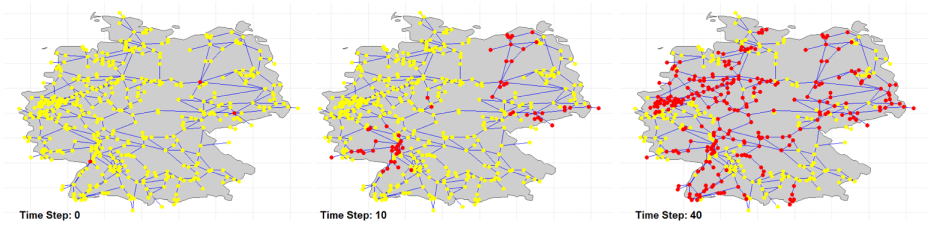}\\
            (c) German power grid
			 \end{tabular}%
		\caption{Spread of node failures in the European power grid network under the KT model type diffusion.}
		\label{fig:KT_EU}
	\end{figure}

\begin{figure*}[!ht]
    \centering
        \includegraphics[width=0.50\textwidth]{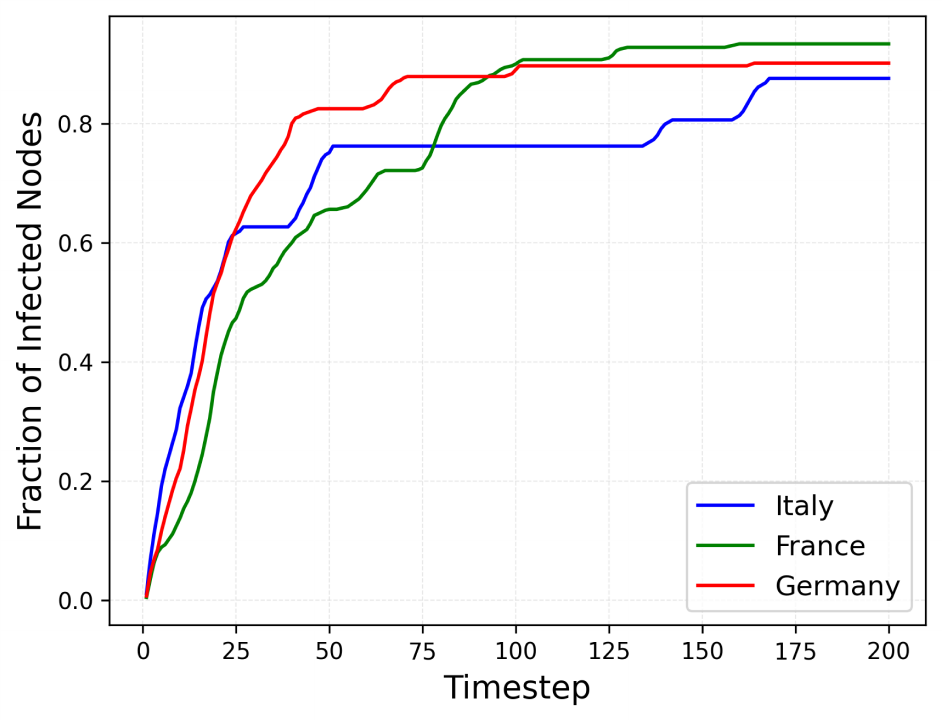}
    \caption{Comparison of fraction of infected nodes over timesteps in 3 European power grid networks.} 
    \label{fig:comparison_deg}
\end{figure*}

Figure~\ref{fig:KT_EU} shows the spread of node failures in the Italian, French, and German networks based on the KT model, where initial infected nodes are selected based on the highest degree centrality. We compute the fraction of infected nodes at different timesteps $t$. Figure~\ref{fig:comparison_deg} shows the fraction of infected nodes in the Italian, French, and German networks. We find that between 1 and 25 timesteps, the diffusion speed in the Italian power grid is higher compared to the German and French power grids, likely due to its smaller size and shorter average path length. Notice that, by timestep 25, 60\% of the nodes in both the Italian and German grids have become infected. From 1 to 75 timesteps, the French power grid exhibits a significantly lower fraction of infected nodes compared to the Italian and German grids. However, after 75 timesteps, the infection spread in the Italian power grid slows down, resulting in a lower fraction of infected nodes than in the German and French grids. This indicates that in the early to mid-stages, diffusion is slower in the French grid, whereas in the long run, the Italian grid experiences the slowest diffusion rate.

To further analyze the differences in diffusion rates among the three power grids, we apply the nonparametric k-sample Anderson-Darling test. The results reveal significant differences in diffusion rates across the Italian, French, and German networks ($p$-value $< 0.001$), highlighting structural variations in these European power grids and their influence on the spread of node failures. Similar patterns are observed when initial infected nodes are selected based on the highest betweenness centrality ($p$-value of the k-sample Anderson-Darling test is less than 0.001, see Online Supplementary Material Figure~S4.).


\subsection{SI model on power grid networks}

In the analysis of the diffusion process in the power grid networks using the SI model, we uniformly assign a constant susceptibility value of 1.0 to every power station $i$, i.e., $\psi_S(i) = 1.0$. Similar to the Ethereum blockchain network, the transmissibility function $\phi_T(i)$ quantifies the overall transmission risk of a power station $j$ by incorporating normalized centrality measures and clustering coefficients as 
$$\psi_T(j)= \phi_1 X_1(j) + \phi_{2} X_{2}(j) + \phi_{3} X_{3}(j),$$
where $X_1$, $X_{2}$, and $X_{2}$ represent normalized degree centrality, normalized betweenness centrality, and normalized clustering coefficient, respectively. The connectivity kernel $\kappa(i, j)$ is employed to account for connections between power stations as 
\[
\kappa(i, j) =
\begin{cases} 
\alpha d_{ij}^{-\gamma}, & \text{if there is a connection between } i \text{ and } j, \\
0, & \text{otherwise}.
\end{cases}
\]
where the distance $d_{ij}$ denotes the Euclidean distance between the geographic coordinates of power stations $i$ and $j$.

The sparks function assigns a constant risk value, denoted as $\epsilon(i) = \zeta$ (Eq.~\eqref{Eq:SI_ILM}), to each power station $i$. This approach provides a comprehensive measure of risk for each station based on its structural significance in the network. We assign weakly informative prior based on the
following distributions

\[
\zeta \sim \text{Exp}(0.0001), \quad \alpha \sim \mathcal{U}(0.001, 5),
\]
\[
\gamma \sim \mathcal{U}(0, 10), \quad \phi_1 \sim \mathcal{U}(0, 5),
\]
\[
\phi_2 \sim \mathcal{U}(0, 5), \quad \phi_3 \sim \mathcal{U}(0, 5).
\]

The Metropolis-Hastings MCMC algorithm runs for a total of 200,000 iterations. The burn-in period was set to 20,000. The computing time taken for the MCMC is about 48 hours on a 32 GB workstation with a 4.40 GHz Intel Core i7 processor.

To assess convergence, we evaluate individual parameter trace plots. The Online Supplementary Material Figure~S6, S7, and S8 show the 
convergences of the model parameter for the power grid of Italy, Germany, and France, respectively. Table~\ref{Bayes_est} shows the posterior mean and 95\% credible intervals of the SI model parameters for three European power grid networks. All the model parameters appear to be statistically significant. The estimated additional risk factors, i.e., the constant risk values ($\hat{\zeta}$) are higher in the German and French power grid. The estimated scale parameters and the exponents of the distance, $\hat{\alpha}$ and $\hat{\gamma}$, respectively, are similar in the three European power grid networks. That is, the spatial distances of the power grid nodes approximately equally affect the diffusion in the Italian, German, and French power grid networks.

We also observe that degree centrality, betweenness centrality, and clustering coefficient are key determinants of transmissibility in the Italian, German, and French power grids. However, the influence of the clustering coefficient is notably stronger in the German power grid compared to the Italian and French power grids, a trend that is also evident in the Ethereum blockchain network (see Table~\ref{Bayes_est2}).

\begin{table*}[!ht]
\centering
\caption{Parameter estimates with 95\% credible intervals for Italy, Germany, and France.}
\label{Bayes_est} 
\resizebox{\textwidth}{!}{
\begin{tabular}{c|cc|cc|cc}
\hline
{Parameter} & \multicolumn{2}{c|}{{Italy}} & \multicolumn{2}{c|}{{Germany}} & \multicolumn{2}{c}{{France}} \\ 
\cline{2-7} 
 & {Mean} & {95\% credible interval} & {Mean} & {95\% credible interval} & {Mean} & {95\% credible interval} \\ 
\hline
 
$\zeta$ & 0.0003 & (0.0001, 0.0003) & 0.0003 & (0.0001, 0.0005) & 0.0002& (0.0001, 0.0003) \\ 
$\alpha$ & 0.0340 & (0.0191, 0.0519) & 0.0313 & (0.0075, 0.0877) & 0.0065 & (0.0039, 0.0114) \\ 
$\gamma$ & 2.2239 & (2.0993, 2.3578) & 2.4890 & (2.2350, 2.6901) & 2.6942 & (2.4689, 2.8499) \\ 
$\phi_1$ & 0.0153 & (0.0009, 0.0522) & 0.8960 & (0.2358, 2.3969) & 3.6950 & (1.8890, 4.9582) \\ 
$\phi_2$ & 2.8538 & (1.5946, 4.3712) & 0.7867 & (0.0814, 2.5497) & 1.4909 & (0.1815, 3.5957) \\ 
$\phi_3$ & 0.1214 & (0.0002, 0.3141) & 3.1224 & (0.7691, 4.8967) & 0.0556 & (0.0012, 0.2199) \\ 
\hline
\end{tabular}
}
\end{table*}


The diffusion of failures within critical infrastructure networks, such as power grids, has significant societal implications, given that these systems are essential to daily life, public safety, and national security. The application of the KT and SI models to European power grids demonstrates that failures can propagate swiftly, influenced by factors such as network topology and node centrality. This high level of interdependence means that localized disruptions, whether due to cyberattacks, adverse weather events, or equipment malfunctions, can escalate into widespread outages, impacting hospitals, transportation systems, communication networks, and emergency services.  The study emphasizes that power grids depend on high-centrality nodes to maintain system coherence. While this structure enhances operational efficiency, it also establishes critical points of failure that may be targeted in cyber-physical attacks. The potential for deliberate disruption poses serious risks to national stability and public confidence. Consequently, infrastructure design must evolve towards more resilient and decentralized architectures, with governments investing in redundancy, real-time monitoring, and robust cybersecurity measures.



\section{Discussion}
\label{Sec:Discussion}
The study provides essential insights into how diffusion processes unfold in complex networks, particularly in cryptocurrency blockchains and critical infrastructure networks. The structural properties of these networks, such as hub-dominated topology, network motifs, and node centrality, significantly impact the speed and extent of diffusion. Understanding these influences is critical for improving network robustness, optimizing information flow, and mitigating vulnerabilities.

We employ epidemic diffusion models, specifically the Kertesz Threshold model and the SI model, to assess the key factors influencing the diffusion process in complex networks. Our simulation study reveals that network structure and topology play a crucial role in shaping diffusion dynamics. Notably, we observe significant variations in diffusion behavior across different network types, including the Erd\H{o}s-R\'{e}nyi network, Geometric Random Graph, and Delaunay Triangulation network. These findings highlight the impact of network architecture on the spread of information or failures within a network.

Furthermore, our findings indicate that network motifs, which are the key measures of network geometry are essential for understanding network diffusion. In blockchain networks, specific motifs can indicate trust relationships or transaction bottlenecks, while in power grid networks, they may highlight structural redundancy and overall robustness. The correlation between diffusion speed and motif concentration offers valuable insights into the underlying mechanisms governing network behavior. Notably, our simulation results reveal that the structural properties of star and square-type motifs enhance the diffusion process by facilitating faster propagation within the network. In contrast, long line-type motifs have a minimal impact on diffusion speed, suggesting that their linear structure does not significantly contribute to the efficiency of failure spread. 

One of the key insights from our case studies is that hub-dominated properties, which are dominant in blockchain networks, provide resilience against random failures but make the network highly vulnerable to targeted attacks. In cryptocurrency blockchains, highly connected hubs facilitate rapid and efficient transaction propagation~\citep{Conoscenti2019,Gebraselase2022,Hanif2023}. However, these hubs also serve as critical points of failure, and their compromise could significantly disrupt network stability. Furthermore, our analysis shows that European power systems exhibit varying degrees of diffusion robustness, highlighting structural differences in their network architectures. Our study also demonstrates that centrality measures, such as degree centrality, betweenness centrality, and clustering coefficient, significantly impact the transmissibility of the SI diffusion process in both blockchain and power grid networks. Additionally, the transaction value between nodes plays a crucial role in exerting infectious pressure within the SI diffusion process in blockchain networks.

These findings have important practical implications, including strengthening cybersecurity in blockchain systems, enhancing the resilience of critical infrastructure networks, and designing more efficient decentralized systems. Future research will focus on exploring adaptive strategies to mitigate targeted attacks, optimizing network structures for controlled diffusion, evaluating cross-network motifs where interdependent networks (e.g., blockchain integrated with financial institutions) influence each other’s diffusion patterns, and developing predictive models that leverage big data and artificial intelligence (AI) to identify vulnerabilities before they lead to cascading failures in interconnected networks.

A key direction for future research is to integrate directionality into the diffusion dynamics. In this study, we treated the Ethereum blockchain and the European power grid as undirected networks to utilize symmetric diffusion models, like the KT and SI models. While this simplification aids our analysis, it overlooks the inherent asymmetry found in real-world systems, such as transaction flows and power distribution. By incorporating directionality, we could develop more accurate models and uncover new insights into the propagation dynamics within complex networks. Future work may also enhance the current analysis by comparing diffusion dynamics between the randomized null models and observed networks, such as cryptocurrency blockchains or critical infrastructure systems.

\section*{Acknowledgements}
The authors thank the associate editor and the two reviewers for their valuable suggestions, which led to significant improvements in the paper.

\section*{Funding}
Asim K. Dey is supported by the National Science Foundation, USA (Grant No. NSF DMS 2453756) and the National Aeronautics and Space Administration, USA (Grant No. 21-AIST21$_{-}$2-0020).

\section*{Data Availability Statement}
All data are publicly available. All R codes are available upon request.

\section*{Conflict of interest}
The authors declare no potential conflict of interest.

\section*{Supplementary materials}
Supplementary files for this preprint are available as Ancillary files on the arXiv abstract page.



\section*{Appendix A}
\label{Appendix_A}
The following algorithm describes the KT diffusion model on networks.
\vspace{0.5cm}
\hrule
\textbf{Algorithm~1: Kertesz Threshold (KT) model }
\hrule    
\begin{algorithmic}[1]
    \Require $\eta(0)$: fraction of nodes that are initially infected, $k$: number of iterations, $\delta$: adopter rate, $\beta$: percentage of blocked nodes, $\tau$: nodes' threshold.
    
    \State $B =$ set of blocked nodes \Comment{Blocked nodes}
    \State $I_t =$ set of infected nodes at timestep $t$ \Comment{Infected nodes}
    \For{$t$ in $\{1,2, \ldots, k\}$}
    \State $I_t = I_{t-1}$
    \For{each $v \in V$}
    \If{$v \notin I_{t-1}$ and $v \notin B$}
    \State $\lambda_v = \frac{|neighbors(v) \cap I_{t-1}|}{|neighbors(v)|}$
    \State $r_v = \text{rand}(0, 1)$ 
    \If{$\lambda_v \geq \tau$ or $r_v \geq \delta$}
    \State add $v$ to $I_t$
    \EndIf
    \EndIf
    \EndFor
    \State \textbf{yield} $I_t$ \Comment{Return iteration status}
    \EndFor
  \end{algorithmic}
\hrule



\section*{Appendix B}
\label{Appendix_B}

\noindent The following procedure was used to generate the mean infection curves and the corresponding 95\% confidence intervals shown in Figure~\ref{fig:three-models}. The confidence intervals are computed using the empirical 2.5th and 97.5th percentiles across 1000 independent simulation runs for each timestep.


\begin{algorithm}
\caption*{\textbf{Algorithm 2:} Estimation of Infection Curves with 95\% Confidence Intervals.}
\begin{algorithmic}[1]

\Require Graph generator $G$ (type, params), number of simulations $N$, timesteps $T$, adopter rate $\delta$, percentage blocked $\beta$, threshold $\tau$
\Ensure Mean infection curve $\mu[1:T]$ and 95\% confidence bounds $\text{CI}_{\text{lower}}[1:T]$, $\text{CI}_{\text{upper}}[1:T]$

\State Initialize matrix $\text{Infection}[1:T, 1:N]$

\For{$i = 1$ to $N$}
    \State $G_i \gets \text{GenerateGraph}(G)$
    \State Seed top 1\% highest-degree nodes as initially infected in $G_i$
    \State Simulate diffusion on $G_i$ using the Kertész Threshold Model (see Algorithm~1) with parameters $\delta$, $\beta$, $\tau$
    \For{$t = 1$ to $T$}
        \State $\text{Infection}[t, i] \gets$ fraction of infected nodes at timestep $t$
    \EndFor
\EndFor

\For{$t = 1$ to $T$}
    \State $\mu[t] \gets \text{mean}(\text{Infection}[t, :])$
    \State $\text{CI}_{\text{lower}}[t] \gets$ empirical 2.5th percentile of $\text{Infection}[t, :]$
    \State $\text{CI}_{\text{upper}}[t] \gets$ empirical 97.5th percentile of $\text{Infection}[t, :]$
\EndFor

\State \Return $\mu[1:T], \text{CI}_{\text{lower}}[1:T], \text{CI}_{\text{upper}}[1:T]$

\end{algorithmic}
\end{algorithm}

\section*{Appendix C}
\vspace*{-2em}  
\label{Appendix_C}

\begin{algorithm}[H]
\caption*{\textbf{Algorithm 3:} Edge-Disjoint 4-Node Motif Detection via WL Hashing.}
\begin{algorithmic}[1]

\Require Graph $G = (V, E)$, motif templates $\mathcal{M} = \{(M_1, h_1), \ldots, (M_6, h_6)\}$
\Ensure Motif counts $\{C_j\}$ for each motif $M_j$, using edge-disjoint instances

\State Initializations: motif counts $C_j \gets 0$ for $j = 1$ to $6$; used edge set $\mathcal{E}_{\text{used}} \gets \emptyset$; 

\ForAll{node $u \in V$}
    \State Let $N(u)$ be the neighbors of $u$
    \Comment{Pass A: Star-expansion}
    \If{$|N(u)| \geq 3$}
        \ForAll{3-combinations $\{v, w, x\} \subset N(u)$}
            \State \Call{ProcessCandidate}{$\{u, v, w, x\}$}
        \EndFor
    \EndIf
    \Comment{Pass B: Path-expansion}
    \ForAll{$v \in N(u)$}
        \ForAll{$w \in N(v)$ such that $w \neq u$}
            \ForAll{$x \in N(w)$ such that $x \notin \{u, v\}$}
                \State \Call{ProcessCandidate}{$\{u, v, w, x\}$}
            \EndFor
        \EndFor
    \EndFor

\EndFor

\Function{ProcessCandidate}{$S$}
    \State $G_S$ is the subgraph induced by $S$; $E_S \gets$ edge set of $G_S$
    \If{$E_S \cap \mathcal{E}_{\text{used}} \neq \emptyset$}
        \State \Return
    \EndIf
    \State Compute WL-hash $h$ of $G_S$
    \ForAll{$(M_j, h_j) \in \text{reverse}(\mathcal{M})$}
        \If{$h = h_j$}
            \State $C_j \gets C_j + 1$; $\mathcal{E}_{\text{used}} \gets \mathcal{E}_{\text{used}} \cup E_S$ 
            \State \textbf{break}
        \EndIf
    \EndFor
\EndFunction

\end{algorithmic}
\end{algorithm}


\bibliographystyle{abbrvnat}
\bibliography{Diffusion_bib,sim,Mobility}

\end{document}